\documentclass[runningheads]{llncs}
\usepackage[T1]{fontenc}
\usepackage{graphicx}
\usepackage{enumitem}
\usepackage{booktabs}
\usepackage{multirow}
\usepackage{tabularx}
\usepackage{makecell} 
\usepackage{amsmath} 
\usepackage{comment}
\usepackage{hyperref}
\usepackage{url}
\usepackage{xurl}
\usepackage{hyperref}

\begin{document}
\title{Security Vulnerabilities in AI-Generated Code: A Large-Scale Analysis of Public GitHub Repositories}
\titlerunning{Security Vulnerabilities in AI-Generated Code}
%
\author{Maximilian Schreiber \and
Pascal Tippe}
\authorrunning{M. Schreiber and P. Tippe}
%
\institute{FernUniversität in Hagen \\ Hagen, Germany \\ \email{maximilian.schreiber@studium.fernuni-hagen.de} \\ \email{pascal.tippe@fernuni-hagen.de}}

\maketitle              
\begin{abstract}

    This paper presents a comprehensive empirical analysis of security vulnerabilities in AI-generated code across public GitHub repositories. We collected and analyzed 7,703 files explicitly attributed to four major AI tools: ChatGPT (91.52\%), GitHub Copilot (7.50\%), Amazon CodeWhisperer (0.52\%), and Tabnine (0.46\%). Using CodeQL static analysis, we identified 4,241 Common Weakness Enumeration (CWE) instances across 77 distinct vulnerability types. Our findings reveal that while 87.9\% of AI-generated code does not contain identifiable CWE-mapped vulnerabilities, significant patterns emerge regarding language-specific vulnerabilities and tool performance. Python consistently exhibited higher vulnerability rates (16.18\%-18.50\%) compared to JavaScript (8.66\%-8.99\%) and TypeScript (2.50\%-7.14\%) across all tools. We observed notable differences in security performance, with GitHub Copilot achieving better security density for Python (1,739 LOC per CWE) and TypeScript, while ChatGPT performed better for JavaScript.
    Additionally, we discovered widespread use of AI tools for documentation generation (39\% of collected files), an understudied application with implications for software maintainability. These findings extend previous work with a significantly larger dataset and provide valuable insights for developing language-specific and context-aware security practices for the responsible integration of AI-generated code into software development workflows.

\end{abstract}
\section{Introduction}

The theoretical foundations of code synthesis date back to Turing's work in the 1940s \cite{copeland2012alan}. However, significant breakthroughs in automated code generation for high-level languages materialized in 2021 through Large Language Models (LLMs) trained on extensive code repositories. AI-based code generation tools have shown significant impact on software development \cite{austin2021programsynthesislargelanguage}. GitHub reports that Copilot improves developer productivity by approximately 55\% and increases developer confidence by up to 85\% \cite{GitHub2023Research}. These statistics highlight the transformative potential of AI code generation. However, they also raise critical questions: Is this confidence justified? What security implications emerge when developers increasingly rely on AI-generated code? Despite growing adoption, systematic analyses of real-world AI-generated code vulnerabilities remain limited. Most existing studies focus on controlled experiments rather than code deployed in production environments. This gap is concerning as these systems become increasingly integrated into development workflows.

To address this research gap, we analyzed AI-generated code from public GitHub repositories attributed to four major tools (ChatGPT, GitHub Copilot, Tabnine, and Amazon CodeWhisperer). We applied static analysis through CodeQL to evaluate the prevalence and patterns of Common Weakness Enumeration (CWE) vulnerabilities in real-world settings. Our study makes several key contributions: (1) A comprehensive analysis of security vulnerabilities across multiple AI code generation tools based on real-world usage, extending previous work with a significantly larger dataset of 7,703 files; (2) Identification of language-specific and tool-specific vulnerability patterns; (3) Analysis of contextual factors influencing security outcomes, including organizational adoption patterns and repository characteristics; and (4) Practical recommendations for mitigating identified risks. The remainder of this paper provides background and reviews related work (Section \ref{sec:back-rel}), details our methodology (Section \ref{sec:meth}), presents results of our analysis (Section \ref{sec:results}), discusses implications with practical recommendations (Section \ref{sec:discussion}), and concludes with Section \ref{sec:conclusion}.
\section{Background and Related Work}\label{sec:back-rel}

AI code generation tools represent a specialized subset of generative AI systems designed to produce syntactically correct and executable source code. These tools are powered by Large Language Models (LLMs) specifically fine-tuned on programming datasets. While general-purpose LLMs primarily generate natural language text, code-specialized models incorporate extensive public repositories of source code during training. They are often supplemented with curated high-quality examples to improve security and correctness. Code generation typically occurs through two primary mechanisms: direct user prompting or contextual auto-completion within integrated development environments (IDEs). These systems also commonly offer auxiliary capabilities such as code documentation generation, refactoring suggestions, and test case creation that all have potential security implications \cite{sarkar2022programmingai}. To analyze the security implications of AI-generated code, we employ several established security frameworks. The Common Weakness Enumeration (CWE) provides a standardized taxonomy for categorizing security vulnerabilities, maintained by the MITRE Corporation. The Common Vulnerabilities and Exposures (CVE) system catalogs specific, publicly disclosed security vulnerabilities with standardized identifiers \cite{MITRE_CWE_FAQ}. For severity assessment, we utilize the Common Vulnerability Scoring System (CVSS) v3.x Base Score, which quantifies vulnerability severity on a scale from 0-10 by evaluating attack vector, complexity, required privileges, user interaction, scope, and impacts to confidentiality, integrity, and availability \cite{FIRST2015CVSS}.

The public release of ChatGPT in late 2022 significantly accelerated both the adoption of AI coding assistants and academic interest in their security implications. Though the body of literature has grown since 2023, this remains an evolving research area with three primary methodological approaches: \textbf{Controlled prompt experiments} evaluate AI tools using predefined programming tasks and analyzing the resulting code for security issues. Hammond et al. \cite{10.1145/3610721} and Khoury et al. \cite{10394237} examined GitHub Copilot and ChatGPT respectively using this approach with CodeQL for static analysis. Yetiştiren et al. \cite{yetiştiren2023evaluatingcodequalityaiassisted} conducted a comparative study across multiple tools (ChatGPT, GitHub Copilot, and Amazon CodeWhisperer) using a similar methodology. \textbf{User studies} as the second methodological approach assess how developers interact with AI code generation tools and the resulting security implications. Perry et al. \cite{10.1145/3576915.3623157} found that participants using Codex introduced more security vulnerabilities while expressing higher confidence in their solutions compared to a control group. However, Sandoval et al. \cite{10.5555/3620237.3620361} reported contradictory findings, with AI-assisted participants demonstrating better security outcomes, highlighting the complexity of human-AI collaboration.

\textbf{Repository mining} analyzes AI-generated code \textit{in the wild} within public repositories. Most similar to our approach, Yujia et al. \cite{10.1145/3716848} examined GitHub repositories for code attributed to GitHub Copilot, employing static analysis tools to identify security vulnerabilities. While their methodology shares similarities with ours, our study significantly expands the scope in three key ways. First, we analyze four major AI tools rather than focusing solely on GitHub Copilot. Second, we examine a substantially larger dataset of 7,703 files. Third, we employ a more systematic approach using a single consistent static analyzer.
\section{Methodology}\label{sec:meth}

Our research methodology follows a systematic approach to investigate security vulnerabilities in AI-generated code across different code generation tools. The process comprises five interconnected stages: (1) selection of representative AI code generation tools based on market adoption and technical capabilities; (2) collection of AI-attributed code samples from public GitHub repositories using the GitHub REST API; (3) identification and application of relevant search terms to ensure accurate attribution of code to specific AI tools; (4) implementation of a multi-stage filtering pipeline to create a clean, analyzable dataset; and (5) static code analysis using CodeQL to identify vulnerabilities and map them to standardized severity metrics through CWE and CVE frameworks.

\subsection{Selection of AI Models}

To select suitable AI code generation tools for our study, we established selection criteria focused on commercially available tools with significant industry adoption. This ensures our findings reflect practical security considerations for contemporary software engineering. \textbf{GitHub Copilot}\footnote{https://github.com/features/copilot}, jointly developed by OpenAI and GitHub, was included as one of the earliest and most widely adopted AI coding assistants. Its GPT-4-based architecture and training on millions of public repositories make it a critical benchmark for examining how large-scale, open-source-derived models handle security concerns. \textbf{ChatGPT}\footnote{https://chat.openai.com}, while not specifically designed for code generation, warrants inclusion due to its widespread application in programming contexts. Despite its general-purpose nature, it has become a popular tool for code and documentation generation. Its inclusion allows us to examine whether general-purpose LLMs potentially lack the security guardrails of dedicated coding tools.

\textbf{Tabnine}\footnote{https://www.tabnine.com} represents a different architectural approach with its emphasis on privacy and configurability, supporting local hosting and multiple LLM backends. Its compliance with licensing norms and selective training data approach may influence vulnerability patterns differently than cloud-only alternatives. \textbf{Amazon CodeWhisperer}\footnote{https://aws.amazon.com/codewhisperer} completes our selection as a major cloud provider's entry in the code generation market. Trained on both open-source and Amazon's proprietary code, it features built-in security scanning functionality, making it particularly relevant for security-focused analysis. These four tools represent different approaches to AI code generation while maintaining significant market presence. We excluded niche or regionally focused tools to maintain methodological clarity, as our GitHub scraping strategy prioritized English-language repositories and globally adopted platforms.

\subsection{Data Collection}

To systematically analyze vulnerabilities in AI-generated code, we leveraged GitHub's extensive repository ecosystem, which hosts over 420 million publicly accessible repositories as of February 2024 \cite{github2025}. The platform's REST API provided a structured mechanism to programmatically identify and collect relevant code artifacts containing references to the AI tools under investigation by querying keywords in code fragments \cite{GitHub2022SearchAPI}. By collecting code from public repositories rather than generating samples in controlled environments, we capture actual vulnerability patterns that emerge when developers integrate AI-generated code into software projects. Our implementation addressed several technical constraints inherent to the GitHub API. To comply with rate-limiting policies, we authenticated requests using a personal access token and implemented timeouts between successive API calls. For queries generating substantial result sets, we navigated the API's pagination mechanism by utilizing the maximum allowable page size of 100 results and systematically traversing all available pages. To circumvent the restriction to 1,000 results per query, we implemented a size-based partitioning strategy that segmented our searches using the size parameter, which controls the permissible file size range for results. For each query approaching the result limit, we manually determined appropriate size ranges through iterative testing to ensure complete coverage across the entire file size spectrum. Our collection was limited to the default branches of repositories, and the GitHub code search doesn't index files larger than 384 KB and may exclude very large repositories \cite{GitHub2022SearchAPI}. Our data collection adhered to GitHub's terms of service and focused exclusively on publicly available repositories to ensure ethical research practices.

\subsection{Identification of Relevant Search Terms}

The GitHub REST API requires a search term for querying code files across repositories. To ensure our analysis captured genuinely AI-generated code, we established two fundamental requirements:

\begin{enumerate}
\item The code must be uniquely attributable to a specific AI code generation tool.
\item The file or code section containing AI-generated code must be clearly identifiable.
\end{enumerate}

These requirements were best satisfied through carefully selected search terms based on attribution patterns in developer comments. Through iterative testing and manual sample verification, we identified six key prefix words that produced the most relevant results when combined with tool names: \textit{by}, \textit{with}, \textit{use}, \textit{used}, \textit{using} and \textit{from}. These prefixes were systematically combined with the names of our four selected AI tools to form complete search terms (e.g., \textit{by+github+copilot} and \textit{using+chatgpt}). The search focused exclusively on English-language terms, reflecting English's dominance in code documentation practices. While developers in non-English speaking regions might use native language attributions, a multilingual approach would introduce methodological challenges including inconsistent translation of tool names and attribution patterns. For tool naming variations, we tested both \textit{Amazon CodeWhisperer} and simply \textit{CodeWhisperer}, finding the latter sufficiently specific. Conversely, we determined that \textit{Copilot} alone appeared in contexts unrelated to the AI tool, so we retained \textit{github+copilot} in our search terms to maintain result quality. We excluded generic terms like \textit{by+ai} or references to underlying models such as \textit{by+GPT} as these would not permit unambiguous attribution to specific tools. We validated our final set of search terms through manual inspection of a sample of results to confirm that they reliably identified genuine AI tool attributions.

\subsection{Data Filtering Pipeline}

The raw data collected from GitHub required substantial processing to create a dataset suitable for vulnerability analysis. We implemented a multi-stage filtering pipeline to systematically refine the collected data. First, we removed duplicates based on the unique combination of filename, SHA1 hash value and search keyword which addresses redundancies created by our size-based partitioning strategy and repository forks. During this stage, we also excluded files containing references to multiple AI tools, as these violated our requirement for unique attribution. Since our research aimed to analyze security vulnerabilities in executable code, we removed non-executable content based on file extensions, filtering out common text file formats such as Markdown, HTML, JSON, and plain text files. After removing non-executable content, we further restricted our dataset to include only files in programming languages supported by our static scanner CodeQL, including C/C++, C\#, Go, Java, JavaScript, Kotlin, Python, Ruby, Swift, and TypeScript. This language selection ensured all retained files could be properly analyzed in subsequent stages. Finally, after manual examination of smaller code samples, we established a minimum threshold of 150 bytes for inclusion, as files below this threshold typically contained minimal executable content such as simple \textit{Hello World} programs or trivial function declarations.

\subsection{Static Code Analysis and Vulnerability Assessment}

To assess the security of source code, we employed static code analysis which examines it without executing it in a runtime environment. This approach checks code for potential errors, security vulnerabilities, and problematic patterns. Using static code analysis represents standard practice in security research and software development \cite{10.1145/3533767.3534380}. We selected GitHub's CodeQL\footnote{https://codeql.github.com/docs/codeql-overview/about-codeql} (release 2.16.3) as our analysis tool since it supports various programming languages, is open source and integrated into GitHub's platform which underscores its widespread adoption. Furthermore, CodeQL provides excellent coverage of Common Weakness Enumeration (CWE) identifiers with 123 distinct CWEs for Python and 170 each for JavaScript and TypeScript. Concretely, we used the \textit{security-and-quality} query suite for CodeQL, which combines security vulnerability detection with checks for code maintainability and reliability that may lead to future security problems. To map code segments to their AI-generated origins, we established two attribution criteria: The AI-generated portion begins at the line containing the search keyword, and the keyword must appear within a code comment. Files not meeting these criteria were excluded from our analysis. For severity assessment, we linked identified CWEs to CVEs through the National Vulnerability Database\footnote{https://nvd.nist.gov} maintained by the U.S. National Institute of Standards and Technology. This process utilized the NVD API to retrieve CVEs associated with detected CWEs. To ensure consistent severity ratings, we exclusively used CVSS v3.X scores (Common Vulnerability Scoring System version 3), which provide standardized risk assessments on a 0-10 scale. This approach inherently excluded older CVEs still using the deprecated CVSS v2 framework, ensuring our analysis reflects current vulnerability prioritization practices \cite{FIRST2015CVSS}.
\section{Results}\label{sec:results}

\subsection{GitHub Repository Search Results}

Our search methodology yielded a substantial dataset of potentially AI-generated code samples from public GitHub repositories. Table~\ref{tab:attribution_patterns} presents the distribution of search results across the four AI code generation tools investigated, broken down by specific search terms. In total, our queries returned 82,413 potential files containing AI-generated code, with significant differences in prevalence observed between tools. We conducted all searches in February 2024 and archived the results for subsequent analysis. The snapshot approach ensures reproducibility of our analysis while acknowledging that the prevalence of AI-attributed code likely continues to increase over time.

\begin{table}[htb]
    \centering
    \setlength{\tabcolsep}{6pt}
    \caption{Distribution of GitHub search results by attribution pattern across AI code generation tools.}
    \label{tab:attribution_patterns}
    \begin{tabular}{lrrrr}
    \toprule
    & \multicolumn{4}{c}{\textbf{AI Code Generation Tools}} \\
    \cmidrule{2-5}
    \multicolumn{1}{c}{\textbf{Attribution}} & \multicolumn{1}{c}{\textbf{ChatGPT}} & \multicolumn{1}{c}{\textbf{GitHub}} & \multicolumn{1}{c}{\textbf{Tabnine}} & \multicolumn{1}{c}{\textbf{[Amazon]}} \\
    \multicolumn{1}{c}{\textbf{pattern}} & & \multicolumn{1}{c}{\textbf{Copilot}} & & \multicolumn{1}{c}{\textbf{CodeWhisperer}} \\
    \midrule
    \textit{by+}    & 21,503 & 2,300 & 67  & 63 \\
    \textit{with+}  & 23,569 & 1,000 & 200 & 248 \\
    \textit{use+}  & 12,866 & 800   & 200 & 61 \\
    \textit{used+}  & 1,900  & 200   & 3   & 12 \\
    \textit{using+} & 7,700  & 900   & 44  & 90 \\
    \textit{from+}  & 8,073  & 500   & 63  & 51 \\
    \midrule
    \textbf{Total} & \textbf{75,611} & \textbf{5,700} & \textbf{577} & \textbf{525} \\
    \bottomrule
    \end{tabular}
\end{table}

The search results reveal notable disparities in the prevalence of attribution comments across different AI code generation tools. ChatGPT demonstrates the highest attribution frequency with 75,611 results (91.7\% of the total), followed by GitHub Copilot with 5,700 results (6.9\%). Tabnine and Amazon CodeWhisperer show considerably lower attribution rates with 577 (0.7\%) and 525 (0.6\%) results respectively. These disparities may reflect differences in market adoption, user practices regarding attribution, or variations in the typical use cases for each tool. It is worth noting that our search methodology captures only explicitly attributed AI-generated code, which likely represents a fraction of the total AI-assisted code in public repositories. Many developers may utilize these tools without including attribution comments, particularly in professional or commercial contexts. Therefore, our dataset represents a conservative estimate of AI tool usage, primarily capturing cases where developers deliberately acknowledged AI assistance.

\subsection{Filtering Results and Dataset Characteristics}

Following the collection of 82,413 potential AI-generated code samples, we applied our multi-stage filtering pipeline as described in the methodology section. Table~\ref{tab:filtering_results} presents a comprehensive overview of each filtering stage and its impact on the dataset. The first filtering stage removed 22,736 duplicate entries, primarily resulting from our size-based partitioning strategy and the presence of forked repositories on GitHub. Additionally, we filtered out 132 files containing attributions to multiple AI code generation tools to maintain our requirement for unique tool attribution. 

This initial stage reduced the dataset by approximately 27.59\%. The text file filtering stage had the most substantial impact, removing 42,615 files (51.71\% of the original dataset). The largest categories of excluded files were Markdown documentation (.md, .markdown: 19,847 files), HTML documents (.html: 10,613 files), JSON data files (.json: 5,134 files), and plain text files (.txt: 3,121 files). Additional excluded formats included data files (.csv: 2,326 files), markup documents (.xml: 677 files), configuration files (.yaml and .yml: 629 files), and typesetting documents (.tex: 268 files). Language compatibility filtering removed an additional 6,634 files written in programming languages not supported by our chosen static analysis tool CodeQL. After applying the minimum file size threshold of 150 bytes, which eliminated 41 trivial code samples, our final dataset comprised 10,387 files suitable for vulnerability analysis.

\begin{table}[htb]
    \centering
    \caption{Results of the data filtering pipeline applied to collected AI-attributed code files.}
    \label{tab:filtering_results}
    \begin{tabular}{lrrr}
    \toprule
    \multicolumn{1}{c}{\textbf{Filtering stage}} & \multicolumn{1}{c}{\textbf{Files}} & \multicolumn{1}{c}{\textbf{Files}} & \multicolumn{1}{c}{\textbf{Reduction}} \\
    & \multicolumn{1}{c}{\textbf{filtered}} & \multicolumn{1}{c}{\textbf{remaining}} & \multicolumn{1}{c}{\textbf{(\%)}} \\
    \midrule
    Raw data collection & - & 82,413 & - \\
    Removal of duplicates & 22,736 & 59,677 & 27.59\% \\
    Removal of text files & 42,615 & 17,062 & 51.71\% \\
    Language compatibility filtering & 6,634 & 10,428 & 8.05\% \\
    File size filtering & 41 & 10,387 & 0.05\% \\
    \midrule
    \textbf{Total reduction} & \textbf{72,026} & \textbf{10,387} & \textbf{87.40\%} \\
    \bottomrule
    \end{tabular}
\end{table}

\begin{table}[htb]
    \centering
    \caption{Distribution of filtered AI-attributed code files by AI code generation tool.}
    \label{tab:ai_tool_distribution}
    \begin{tabular}{lrr}
    \toprule
    \textbf{AI Tool} & \textbf{File count} & \textbf{Percentage} \\
    \midrule
    ChatGPT & 9,506 & 91.52\% \\
    GitHub Copilot & 779 & 7.50\% \\
    Amazon CodeWhisperer & 54 & 0.52\% \\
    Tabnine & 48 & 0.46\% \\
    \midrule
    \textbf{Total} & \textbf{10,387} & \textbf{100.00\%} \\
    \bottomrule
    \end{tabular}
\end{table}

Table~\ref{tab:ai_tool_distribution} displays the distribution of files across different AI code generation tools in our filtered dataset. This distribution offers valuable insights into the relative adoption rates of these tools within public GitHub repositories. ChatGPT remains overwhelmingly dominant, representing 91.52\% of the filtered dataset, followed by GitHub Copilot at 7.50\%. Amazon CodeWhisperer and Tabnine account for only 0.52\% and 0.46\% of the dataset, respectively. These proportions closely mirror the distribution observed in our raw dataset, suggesting that our filtering process did not introduce significant bias with respect to AI tool representation.

\begin{table}[htb]
    \centering
    \caption{Programming language distribution across AI code generation tools.}
    \label{tab:tool_language_distribution}
    \resizebox{\textwidth}{!}{
    \begin{tabular}{lrrrrrrrr|rr}
    \toprule

    \textbf{Programming} & \multicolumn{2}{c}{\textbf{ChatGPT}} & \multicolumn{2}{c}{\textbf{GitHub Copilot}} & \multicolumn{2}{c}{\textbf{Amazon}} & \multicolumn{2}{c}{\textbf{Tabnine}} & \multicolumn{2}{c}{\textbf{Total}}

    \\
    \multicolumn{1}{c}{\textbf{language}} & \multicolumn{2}{c}{} & \multicolumn{2}{c}{\textbf{Copilot}} & \multicolumn{2}{c}{\textbf{CodeWhisperer}} & \multicolumn{2}{c}{} & \multicolumn{2}{c}{} \\
    \cmidrule(lr){2-3} \cmidrule(lr){4-5} \cmidrule(lr){6-7} \cmidrule(lr){8-9} \cmidrule(lr){10-11}
    & \textbf{Count} & \textbf{\%} & \textbf{Count} & \textbf{\%} & \textbf{Count} & \textbf{\%} & \textbf{Count} & \textbf{\%} & \textbf{Count} & \textbf{\%} \\
    \midrule
    Python & 3,801 & 39.99\% & 175 & 22.46\% & - & - & 12 & 25.00\% & 3,988 & 38.34\% \\
    JavaScript & 2,029 & 21.34\% & 90 & 11.55\% & 2 & 3.70\% & 3 & 6.25\% & 2,124 & 20.45\% \\
    TypeScript & 1,485 & 15.62\% & 69 & 8.86\% & 28 & 51.85\% & 15 & 31.25\% & 1,597 & 15.38\% \\
    C\# & 622 & 6.54\% & 28 & 3.59\% & - & - & - & - & 650 & 6.26\% \\
    C/C++ & 533 & 5.61\% & 311 & 39.92\% & 4 & 7.41\% & 4 & 8.33\% & 852 & 8.18\% \\
    Java & 533 & 5.61\% & 84 & 10.78\% & - & - & 16 & 33.33\% & 633 & 6.09\% \\
    Go & 243 & 2.56\% & 12 & 1.54\% & 3 & 5.56\% & - & - & 258 & 2.48\% \\
    Kotlin & 97 & 1.02\% & 6 & 0.77\% & 17 & 31.48\% & - & - & 120 & 1.21\% \\
    Swift & 91 & 0.96\% & 1 & 0.13\% & - & - & - & - & 92 & 0.89\% \\
    Ruby & 72 & 0.76\% & 3 & 0.39\% & - & - & - & - & 75 & 0.72\% \\
    \midrule
    \textbf{Total} & \textbf{9,506} & \textbf{100\%} & \textbf{779} & \textbf{100\%} & \textbf{54} & \textbf{100\%} & \textbf{48} & \textbf{100\%} & \textbf{10,387} & \textbf{100\%} \\
    \bottomrule
    \end{tabular}
    }
\end{table}

Table~\ref{tab:tool_language_distribution} presents a detailed breakdown of programming languages across different AI code generation tools in our filtered dataset. The dominance of Python (38.34\% overall) is particularly noteworthy and aligns with the fact that Python constitutes a significant portion of the training data used by various AI code generation tools, particularly GitHub Copilot. JavaScript (20.45\%) and TypeScript (15.38\%) also show strong representation, reflecting their widespread use in web development. The overall language distribution correlates reasonably well with GitHub's 2022 language popularity statistics, with some notable exceptions \cite{GitHub2022Programming}. Command-line languages such as Shell and PowerShell are substantially underrepresented in our dataset, suggesting either lower usage of AI tools for scripting tasks or different attribution patterns in these contexts. Additionally, Java appears somewhat underrepresented compared to its general popularity on GitHub, which may indicate differences in how developers utilize AI assistance across programming paradigms.

Analysis of language distribution across individual AI tools reveals distinct specialization patterns. While Python dominates ChatGPT's output (39.99\%), C/C++ represents the largest category for GitHub Copilot (39.92\%). Amazon CodeWhisperer shows notable specialization in TypeScript (51.85\%) and Kotlin (31.48\%), while Tabnine demonstrates strength in Java (33.33\%) and TypeScript (31.25\%). These distribution patterns are not uniform, indicating that different AI code generation tools possess varying degrees of specialization or exhibit uneven usage intensities across programming languages.

\subsection{CodeQL Analysis}

\begin{table}[htb]
    \centering
    \setlength{\tabcolsep}{4pt} 
    \caption{CodeQL analysis metrics across AI tools by programming language.}
    \label{tab:codeql_metrics}
    \begin{tabular}{llrrr}
    \toprule
    \textbf{AI Tool} & \textbf{Language} & \textbf{Files} & \textbf{Erroneous files} & \textbf{LOC} \\
    \midrule
    \multirow{3}{*}{ChatGPT} & TypeScript & 1,485 & 3 & 224,468 \\
    & JavaScript & 2,029 & 20 & 371,251 \\
    & Python & 3,794 & 49 & 565,991 \\
    \midrule
    \multirow{2}{*}{\shortstack[l]{Amazon\\ CodeWhisperer}} & TypeScript & 28 & 0 & 6,525 \\
    & JavaScript & 2 & 0 & 59 \\
    \midrule
    \multirow{3}{*}{GitHub Copilot} & TypeScript & 69 & 41 & 13,187 \\
    & JavaScript & 90 & 1 & 12,817 \\
    & Python & 175 & 2 & 38,697 \\
    \midrule
    \multirow{3}{*}{Tabnine} & TypeScript & 15 & 0 & 1,857 \\
    & JavaScript & 3 & 0 & 232 \\
    & Python & 6 & 0 & 1,641 \\
    \midrule
    \textbf{Total} & & \textbf{7,696} & \textbf{116} & \textbf{1,236,725} \\
    \bottomrule
    \end{tabular}
\end{table}

As shown in Table \ref{tab:tool_language_distribution}, our dataset includes 10 different programming languages. We further limited the results to three programming languages: Python, JavaScript, and TypeScript. These languages were selected because they collectively represent 74\% of the dataset and are typically used in web and application development contexts where security vulnerabilities often have critical consequences. The interpretable nature of these languages eliminates compilation requirements and prevent errors from potentially uncompilable code fragments in repositories. From the remaining 7,703 files, we excluded 586 files that contained the search keywords in non-comment contexts such as console logs or exception blocks, further reducing the dataset to 7,117 analyzable files. Notably, 47.8\% of exclusions (280 files) stemmed from false positives where \textit{from} appeared in import statements rather than attribution comments.

Subsequently, we used CodeQL to systematically examine the collected code.  Table \ref{tab:codeql_metrics} presents statistics for these eleven databases, with data sourced from the diagnostic and metric output generated by the CodeQL CLI during analysis. In total, we analyzed 7,696 files comprising 1,236,725 lines of code across four AI code generation tools and three programming languages. ChatGPT-generated code constituted the majority of our dataset, with 3,794 Python files (565,991 lines of code), 2,029 JavaScript files (371,251 lines), and 1,485 TypeScript files (224,468 lines). GitHub Copilot contributed 334 files (64,701 lines), while Amazon CodeWhisperer and Tabnine had significantly smaller representations with 30 files (6,584 lines) and 24 files (3,730 lines), respectively. During our analyses, CodeQL identified syntax errors in 116 files (approximately 1.5\% of the total dataset). These erroneous files were distributed across AI tools and languages, with GitHub Copilot-generated TypeScript files showing the highest error rate (41 of 69 files). However, the overall error rate remains negligibly low and is unlikely to significantly impact our analysis results.

Subsequent CodeQL analysis of 7,696 files (1,236,725 LOC) revealed ChatGPT-generated code constitutes 91.4\% of the dataset, with 3,794 Python files (565,991 LOC), 2,029 JavaScript files (371,251 LOC), and 1,485 TypeScript files (224,468 LOC). Smaller contributions came from GitHub Copilot (334 files), Amazon CodeWhisperer (30 files), and Tabnine (24 files), reflecting real-world adoption patterns. Syntax errors affected only 1.5\% of files (116), primarily in GitHub Copilot's TypeScript files (41 errors), though this localized issue doesn't invalidate broader trends. All results were exported in CSV/SARIF formats for reproducibility. Our CodeQL query suite categorized findings into three severity levels:

\begin{itemize}
    \item \textbf{Error}: Critical security flaws requiring immediate remediation
    \item \textbf{Warning}: Potential vulnerabilities needing review
    \item \textbf{Recommendation}: Code quality improvements (e.g., unused variables)
\end{itemize}

Security-relevant findings (errors/warnings) constituted 36.8\% of total alerts (5,892/16,308), with recommendations dominating at 63.2\%. While 46.4\% of files (3,568) contained findings, 53.6\% (4,128 files) passed static analysis entirely without any finding. However, 25.1\% of files (1,932) had only recommendations, suggesting technical debt accumulation through unused code (63.1\% of recommendations) and commented-out code fragments (8.1\%).

\subsection{Vulnerability Analysis}

To contextualize CodeQL findings within established vulnerability frameworks, we mapped CodeQL findings to CWEs by cross-referencing raw CWE metadata from MITRE's official CSV files. This process excluded non-security-related CodeQL alerts (e.g., code style recommendations), while focusing only on findings with direct CWE mappings. From 7,117 analyzable files, 861 files (12.1\%) contained at least one CWE-mapped vulnerability, resulting in 4,241 distinct CWE occurences. 

As shown in Table \ref{tab:merged_cwe_analysis}, ChatGPT-generated code accounted for 94.9\% of vulnerable files (817/861) and 92.9\% of CWEs (3,943/4,241), while GitHub Copilot contributed 4.9\% of files (42/861) and 6.9\% of CWEs (295/4,241). Tabnine showed minimal impact (2 files, 3 CWEs), and Amazon CodeWhisperer exhibited no vulnerabilities. Notably, 87.9\% of analyzed files were free from CWE vulnerabilities, indicating that the majority of AI-generated code in our dataset does not contain detectable security weaknesses. The absence of CWE-mapped vulnerabilities in Amazon CodeWhisperer-generated files is particularly interesting. Similarly, Tabnine demonstrated strong security performance with only two files containing vulnerabilities. 

However, this conclusion must be interpreted cautiously given the significantly smaller sample sizes for these tools compared to ChatGPT and GitHub Copilot. Across the 861 files containing vulnerabilities, we identified 77 distinct types of CWEs, representing a broad spectrum of security weaknesses. The distribution of these vulnerabilities reveals interesting patterns when analyzed by programming language. Python code consistently exhibited higher vulnerability rates (16.18\% to 18.50\%) compared to JavaScript (8.66\% to 8.99\%) and TypeScript (2.50\% to 7.14\%). This language-dependent pattern persisted across AI tools, suggesting that the vulnerability profile is more strongly influenced by programming language characteristics than by the specific AI system generating the code.

\begin{table}[htb]
    \centering
    \caption{CWE vulnerability distribution by AI tool and programming language.}
    \label{tab:merged_cwe_analysis}
    \begin{tabular}{llrrrr}
    \toprule
    \multirow{2}{*}{\textbf{AI Tool}} & \multirow{2}{*}{\textbf{Language}} & \multicolumn{3}{c}{\textbf{File Analysis}} & \textbf{Code Density} \\
    \cmidrule(lr){3-5}
    & & \textbf{Files} & \textbf{CWEs} & \textbf{Prevalence (\%)} & \textbf{LOC per CWE} \\
    \midrule
    \multirow{3}{*}{ChatGPT} 
    & Python & 606 & 2,468 & 16.18 & 399 \\
    & JavaScript & 174 & 1,371 & 8.66 & 932 \\
    & TypeScript & 37 & 104 & 2.50 & 444 \\
    \midrule
    \multirow{3}{*}{GitHub Copilot} 
    & Python & 32 & 238 & 18.50 & 1,739 \\
    & JavaScript & 8 & 40 & 8.99 & 393 \\
    & TypeScript & 2 & 17 & 7.14 & 905 \\
    \midrule
    \multirow{2}{*}{Tabnine} 
    & Python & 1 & 2 & 16.67 & 686 \\
    & TypeScript & 1 & 1 & 6.67 & 54 \\
    \midrule
    \multirow{2}{*}{\shortstack[l]{Amazon\\CodeWhisperer}} & \multirow{2}{*}{--} & \multirow{2}{*}{0} & \multirow{2}{*}{0} & \multirow{2}{*}{0.00} & \multirow{2}{*}{--} \\
    & \\
    \midrule
    \multicolumn{2}{l}{\textbf{Overall Average}} & 861 & 4,241 & 11.36 & 650 \\
    \bottomrule
    \end{tabular}
\end{table}

To better understand the density of vulnerabilities in the generated code, we calculated the average lines of code per CWE for each AI tool and programming language combination. This metric represents the average number of code lines that can be generated before encountering a security vulnerability, with higher values indicating better security performance. As highlighted in Table \ref{tab:merged_cwe_analysis}, GitHub Copilot achieved the best security density for Python (1,739 LOC per CWE) and TypeScript (905 LOC per CWE), while ChatGPT performed best for JavaScript (932 LOC per CWE). These substantial differences in vulnerability density between tools are particularly noteworthy because they contrast with the relatively consistent file-level prevalence rates observed within each programming language. The similarity in vulnerability prevalence between TypeScript and JavaScript (particularly evident in GitHub Copilot's output at 7.14\% and 8.99\% respectively) likely stems from their close relationship as programming languages. Additionally, CodeQL uses the same query suite for both languages, resulting in identical CWE coverage profiles.

\subsection{Distribution and Severity of CWE Types}

To provide deeper insights into vulnerability patterns, we analyzed the distribution of specific CWE types across AI tools and programming languages. Across all 861 vulnerable files, we identified 77 distinct CWE types, with pronounced differences in their distribution patterns. Table \ref{tab:cwe_distribution} in Appendix \ref{appendix:table} details the distribution of top CWEs by programming language across AI Tools and shows both consistent and tool-specific vulnerability patterns. Of particular significance in our analysis are the differences between the analyzed AI tools across programming languages. In Python code generated by ChatGPT, CWE-772 (Missing Release of Resource after Effective Lifetime) accounts for approximately 5.75\% of vulnerabilities, while this CWE does not appear among GitHub Copilot's top five vulnerabilities for the same language. For JavaScript, GitHub Copilot frequently generates code containing CWE-676 (Use of Potentially Dangerous Functions), representing 35.00\% of its JavaScript vulnerabilities, while this weakness does not appear in ChatGPT's top five for JavaScript. In TypeScript, the relatively critical CWE-020 (Improper Input Validation) appears in ChatGPT's output (12.50\%) but is absent from GitHub Copilot's top vulnerabilities in this language.

These differential patterns suggest distinct security characteristics in the code generation mechanisms of these AI tools. The mean distance (in lines of code) between the first security-relevant vulnerability and the nearest preceding code comment containing our search terms was approximately 121.11 lines, suggesting that vulnerabilities typically appear well after the attribution point. However, this arithmetic mean has limited informative value due to the high coefficient of variation (2.52) calculated from the data. The median distance was only 43 lines of code which is substantially lower than the arithmetic mean and demonstrates the presence of extreme outliers in the dataset. 

To assess the severity of identified vulnerabilities, we analyzed 62,220 National Vulnerability Database (NVD) entries that corresponded to 64 distinct CWEs. Notably, 13 CWEs identified by CodeQL had no corresponding CVEs in the NVD, which we excluded from our severity calculations. To avoid skewing results with rarely occurring vulnerabilities, we only included CWEs associated with more than ten CVEs. Average CVSS Base Scores for each CWE were calculated using the arithmetic mean of all associated CVEs' Base Scores. For this particular metric, the arithmetic mean provides meaningful insight as the average coefficient of variation across CWEs was approximately 0.1749 which indicates relatively consistent severity ratings within each CWE category. The five most critical CWEs identified in our dataset based on average CVSS Base Scores are detailed in Table  \ref{tab:critical_cwes}. It is noteworthy that four of these five CWEs (with the exception of CWE-259) appear in MITRE's 2024 Top 25 Most Dangerous Software Weaknesses list \cite{MITRE2024CWETop25}. Additionally, the severity scores of these top five CWEs fall within a narrow range, with a maximum difference of only 0.19 points, indicating comparable criticality levels among these vulnerability types.

\begin{table}[htb]
\centering
\caption{Most critical CWEs ranked by average CVSS base score.}
\label{tab:critical_cwes}
\begin{tabular}{llc}
\toprule
\textbf{CWE-ID} & \textbf{CWE description} & \textbf{Avg. CVSS score} \\
\midrule
CWE-89 & SQL Injection & 8.76 \\
CWE-78 & OS Command Injection & 8.68 \\
CWE-94 & Code Injection & 8.68 \\
CWE-259 & Use of Hard-coded Password & 8.64 \\
CWE-798 & Use of Hard-coded Credentials & 8.57 \\
\bottomrule
\end{tabular}
\end{table}

\section{Discussion}\label{sec:discussion}

\subsection{Main Findings}

Our systematic analysis of AI-generated code in public GitHub repositories reveals several significant patterns with important security implications:

\textbf{Code security prevalence.} The majority (87.9\%) of AI-generated code lacks identifiable CWE-mapped vulnerabilities. When examining all CodeQL findings, 53.68\% of files triggered no alerts whatsoever, while 25.08\% contained only minor recommendations which are primarily unused code elements (63.12\%) and commented-out code (8.08\%).

\textbf{Documentation generation.} We discovered significant AI tool usage for documentation purposes, with 39\% (23,236) of collected deduplicated files being documentation formats (.md, .markdown, .txt, .tex). 8,320 filenames containing \textit{readme} confirm this widespread practice and suggest AI tools serve dual purposes in development workflows with enormous impact on software maintainability.

\textbf{Tool adoption patterns.} ChatGPT dominates our dataset (91.52\%), with GitHub Copilot (7.5\%). Amazon CodeWhisperer (0.52\%) and Tabnine (0.46\%) represent smaller portions which reflect current adoption in public repositories. This distribution pattern suggests that general-purpose LLMs are more widely used for code generation than specialized coding assistants in public contexts

\textbf{Language-specific vulnerability profiles.} Python consistently exhibited higher vulnerability rates (16.18\%-18.50\%) compared to JavaScript (8.66\%-8.99\%) and TypeScript (2.50\%-7.14\%) across all AI tools. This suggests a stronger influence on vulnerabilities by language characteristics than by the AI system itself.

\textbf{Security density variation.} GitHub Copilot achieved better security performance for Python (1,739 LOC per CWE) and TypeScript (905 LOC per CWE), while ChatGPT performed better for JavaScript (932 LOC per CWE), indicating tool-specific strengths across different languages.

\textbf{Vulnerability pattern distribution.} We identified both consistent and tool-specific vulnerability patterns across 77 CWE types. Some vulnerabilities appeared consistently within specific languages, while others were unique to particular AI systems, such as CWE-772 appearing frequently in ChatGPT's Python code but not in GitHub Copilot's output.

\textbf{Critical vulnerability types.} Five CWEs had particularly high average CVSS Base Scores: SQL Injection (CWE-89), OS Command Injection (CWE-78), Code Injection (CWE-94), and hard-coded credentials (CWE-259/798). Four of these appear in MITRE's 2024 Top 25 Most Dangerous Software Weaknesses list \cite{MITRE2024CWETop25} and indicate that AI systems still generate code with widely-known and severe security weaknesses.

\subsection{Practical Implications}

The substantial variation in vulnerability patterns across programming languages suggests that security considerations should be tailored to specific language contexts. Python's consistently higher vulnerability rates (16.18\%-18.50\%) across all tools indicate that teams working primarily with this language should implement more rigorous security controls when incorporating AI-generated code. Our identification of dominant CWE types within each language provides a foundation for targeted code review procedures focused on language-specific vulnerability patterns. Security teams should prioritize reviewing for resource management issues in Python, potentially dangerous function calls in JavaScript, and improper input validation in TypeScript based on our observed patterns. The differential security performance of various AI tools across programming languages presents both challenges and opportunities for organizations. Our findings suggest that no single tool provides optimal security across all contexts, with GitHub Copilot demonstrating better security density for Python (1,739 LOC per CWE) and TypeScript (905 LOC per CWE), while ChatGPT performed better for JavaScript (932 LOC per CWE). 

Organizations might benefit from a strategic approach that selects specific AI tools for particular programming tasks based on their security characteristics in those contexts, rather than adopting a single solution. For instance, our findings suggest that GitHub Copilot might be preferable for Python and TypeScript development when security is a priority, while ChatGPT could be more appropriate for JavaScript tasks. The relatively high prevalence of code quality issues in AI-generated code suggests that while these tools can rapidly generate functional code, they may introduce maintenance challenges through unnecessary complexity and technical debt. Development teams should implement processes to review and refactor AI-generated code specifically targeting these common quality issues. Our finding that 25.08\% of files contained only recommendations (primarily code quality issues) while having no security-critical findings underscores the importance of addressing these maintenance concerns separately from security reviews. Additional contextual factors may influence the security of AI-generated code beyond tool-specific and language-specific considerations. For a detailed analysis of organizational adoption patterns across different sectors and the relationship between repository popularity metrics and vulnerability characteristics, see Appendix \ref{appendix:contextual}.

\subsection{Limitations}

Our methodology presents several constraints that should be considered when interpreting our findings. The reliance on explicit attribution comments to identify AI-generated code likely underrepresents actual AI tool usage, as developers may not consistently include such attributions. This selection bias potentially skews our dataset toward code from educational contexts, less experienced developers, or repositories with more rigorous documentation practices. The substantial disparity in sample sizes between tools (ChatGPT at 91.52\% versus Amazon CodeWhisperer at 0.52\% and Tabnine at 0.46\%) limits the statistical power of comparative analyses and warrants cautious interpretation of conclusions regarding less-represented tools. Still, these percentages accurately reflect real-world adoption patterns. Our static analysis approach using CodeQL introduces inherent limitations, as it cannot identify runtime issues, logical flaws, or security weaknesses that manifest only during execution. Furthermore, the identified relationships between tool usage and vulnerability patterns do not necessarily imply causation. Confounding factors such as developer experience, project complexity, development practices, and external code review processes may simultaneously influence both tool selection and code security outcomes. The focus on English-language search terms represents another methodological constraint. While English dominates as the primary language for code documentation on GitHub, developers in non-English speaking regions might attribute AI assistance in their native languages. However, implementing a multilingual search approach would introduce significant methodological challenges, including varying attribution patterns across languages, inconsistent translation of tool names, and difficulties ensuring balanced representation across language communities. Finally, our findings represent repository snapshots at a specific point in time and do not account for the evolution of AI tools or changing developer interactions with these systems. The rapid advancement in AI code generation capabilities suggests newer models may exhibit different vulnerability patterns than those captured in our dataset.

\subsection{Future Work}

Several promising research directions emerge from our findings and limitations. Longitudinal studies tracking vulnerability patterns over time could provide valuable insights into how AI code generation tools evolve and whether security characteristics improve with model advancements. Such research could also examine whether developers become more adept at using these tools securely as they gain experience with AI-assisted programming. Controlled experiments comparing human-written code against AI-generated alternatives for identical tasks could help isolate the causal factors behind security weaknesses. This approach would complement our observational methodology by providing more definitive evidence regarding the security implications of AI code generation tools under controlled conditions. As new models and tools emerge, comparative studies of their security characteristics will help developers make informed choices. Particularly valuable would be examining how models specifically trained with security awareness compare to general-purpose code generation systems in terms of vulnerability introduction rates and patterns. Research into effective organizational policies and practices for secure integration of AI-generated code into production systems would provide practical guidance for industry adoption. This could explore training requirements, review procedures, and governance frameworks specific to AI-assisted development. Additionally, investigating tool-specific mitigation strategies based on the vulnerability patterns we identified could lead to practical improvements in secure AI-assisted development practices.
\section{Conclusion}\label{sec:conclusion}

Generative AI is transforming software development through automated code generation and documentation creation. Our analysis of AI-generated code in public GitHub repositories reveals that while most code files (87.9\%) does not contain identifiable CWE-mapped vulnerabilities, relevant patterns still emerged that warrant attention from developers and security teams. We found substantial differences in vulnerability profiles across programming languages with Python consistently exhibiting higher vulnerability rates than JavaScript and TypeScript. This suggests that security considerations should be language-specific. Similarly, variations in security performance between tools (GitHub Copilot performing better for Python and TypeScript, ChatGPT for JavaScript) indicate no single tool provides optimal security across all contexts. Our discovery that AI tools are widely used for documentation purposes (39\% of collected files) highlights an additional use case with important implications for software maintainability. As these tools become increasingly integrated into development workflows, developers must approach them with appropriate caution. By understanding the specific vulnerability patterns identified in our study, organizations can develop targeted security practices to mitigate risks while capitalizing on the transformative potential of AI-assisted software development.

%
%
\begin{credits}

    \subsubsection{\discintname}
        The authors have no competing interests to declare that are relevant to the content of this article.
\end{credits}
\newpage
\appendix
\section{Contextual Analysis}\label{appendix:contextual}

To understand organizational patterns in AI tool adoption, we analyzed committer email domains from our dataset, categorizing them into academic (.edu or containing \textit{student}), corporate (.com, .io), government (.gov) and private user groups based on domain suffix patterns. We extracted 1,844,928 email addresses spanning 5,186 unique domains. The classification process involved automated categorization of common domains, followed by manual review of corporate domains to ensure accurate classification. Domains with fewer than 100 occurrences that didn't fall into our predeinfed categories were categorized as unknown/other. Our findings indicate distinct preferences across different organizational types. The distribution analysis revealed that private email addresses dominated our dataset at 76.83\% (1,417,421 addresses), followed by unknown/other at 13.60\% (250,787), corporate addresses at 6.11\% (112,691), university addresses at 3.46\% (63,868), and government addresses at just 0.009\% (161). When examining tool preferences across sectors, we found significant variations that deviate from the overall distribution pattern. Academic institutions showed a stronger tendency toward ChatGPT usage (95.05\%) compared to the dataset average (94.96\%), while using specialized tools like Tabnine and Amazon CodeWhisperer less frequently (0.13\% each compared to the dataset averages of 0.31\% and 0.39\%, respectively). This suggests that university users gravitate toward more general-purpose, widely accessible AI solutions. While ChatGPT remained the most used tool at 83.80\% for corporate entities, this was substantially lower than the dataset average of 94.96\%. Most notably, specialized code generation tools were drastically overrepresented in corporate environments, with Tabnine usage approximately 10 times higher (3.12\% vs. 0.31\%) and Amazon CodeWhisperer about 22 times more prevalent (8.41\% vs. 0.39\%) than in the general dataset.  The predominance of ChatGPT in academic settings may be attributed to its free accessibility and widespread familiarity, while GitHub Copilot's slightly elevated usage among university emails (4.69\% vs. 4.34\% dataset average) may reflect the availability of free licenses through GitHub's Education Program. In corporate environments, factors beyond cost and familiarity appear to drive tool selection, including data privacy concerns, intellectual property considerations, and compatibility with existing enterprise infrastructure.

Beyond organizational adoption patterns, we also examined how repository popularity metrics correlate with security characteristics. We collected stargazer data (user \textit{starring} repositories) from 2,315 projects, totaling 738,476 distinct stargazer entries. Notably, 3,611 repositories (approximately 61\% of all collected repositories) had no stargazers whatsoever but accounted for only 50.3\% of identified CWEs. This is counterintuitive since popular projects should receive more attention and security scrutiny. Only very popular projects with 1,001 or more stargazers have lower CWE rates. Furthermore, the mean CVSS Base Score decreases from 6.73 for repositories with a single stargazer to 5.55 for repositories with 8-100 stargazers, before slightly increasing again for the most popular repositories.

\newpage

\section{Tool-Specific and Language-Specific CWE Vulnerability Patterns}\label{appendix:table}

\begin{table}[h!]
    \centering
    \caption{Distribution of Top CWEs by Programming Language Across AI Tools}
    \label{tab:cwe_distribution}
    \scalebox{0.91}{
    \begin{tabularx}{\textwidth}{l c r X} 
    \toprule
    \textbf{Language} & \textbf{CWE-ID} & \textbf{Count (\%)} & \textbf{MITRE CWE Name} \\
    \midrule
    \multicolumn{4}{l}{\textbf{ChatGPT}} \\
    \midrule
    \multirow{5}{*}{Python} 
    & CWE-563 & 1,044 (42.30\%) & Assignment to Variable without Use \\
    & CWE-396 & 355 (14.38\%) & Declaration of Catch for Generic Exception \\
    & CWE-561 & 184 (7.46\%) & Dead Code \\
    & CWE-772 & 142 (5.75\%) & \makecell[l]{Missing Release of Resource\\after Effective Lifetime} \\
    & CWE-390 & 92 (5.75\%) & Detection of Error Condition Without Action \\
    \midrule
    \multirow{5}{*}{JavaScript} 
    & CWE-563 & 217 (15.83\%) & Assignment to Variable without Use \\
    & CWE-400 & 96 (7.00\%) & Uncontrolled Resource Consumption \\
    & CWE-307 & 81 (5.91\%) & \makecell[l]{Improper Restriction of\\Excessive Authentication Attempts} \\ 
    & CWE-770 & 81 (5.91\%) & Allocation of Resources Without Limits or Throttling \\
    & CWE-570 & 59 (4.30\%) & Expression is Always False \\
    \midrule
    \multirow{5}{*}{TypeScript} 
    & CWE-563 & 13 (12.50\%) & Assignment to Variable without Use \\
    & CWE-020 & 13 (12.50\%) & Improper Input Validation \\
    & CWE-117 & 10 (9.62\%) & Improper Output Neutralization for Logs \\
    & CWE-116 & 9 (8.65\%) & Improper Encoding or Escaping of Output \\
    & CWE-570 & 7 (6.73\%) & Expression is Always False \\
    \midrule
    \multicolumn{4}{l}{\textbf{GitHub Copilot}} \\
    \midrule
    \multirow{5}{*}{Python} 
    & CWE-563 & 136 (57.14\%) & Assignment to Variable without Use \\
    & CWE-390 & 27 (11.34\%) & Detection of Error Condition Without Action \\
    & CWE-396 & 24 (10.08\%) & Declaration of Catch for Generic Exception \\
    & CWE-561 & 19 (7.98\%) & Dead Code \\
    & CWE-685 & 7 (2.94\%) & \makecell[l]{Function Call With Incorrect\\Number of Arguments} \\
    \midrule
    \multirow{5}{*}{JavaScript} 
    & CWE-676 & 14 (35.00\%) & Use of Potentially Dangerous Function \\
    & CWE-570 & 5 (12.50\%) & Expression is Always False \\
    & CWE-571 & 5 (12.50\%) & Expression is Always True \\
    & CWE-020 & 3 (7.50\%) & Improper Input Validation \\
    & CWE-078 & 3 (7.50\%) & OS Command Injection \\
    \midrule
    \multirow{5}{*}{TypeScript} 
    & CWE-400 & 4 (23.53\%) & Uncontrolled Resource Consumption \\
    & CWE-730 & 4 (23.53\%) & Denial of Service \\
    & CWE-1333 & 4 (23.53\%) & Inefficient Regular Expression Complexity \\
    & CWE-570 & 2 (11.76\%) & Expression is Always False \\
    & CWE-571 & 2 (11.76\%) & Expression is Always True \\
    \midrule
    \multicolumn{4}{l}{\textbf{Tabnine}} \\
    \midrule
    Python & CWE-563 & 2 (100.00\%) & Assignment to Variable without Use \\
    \midrule
    TypeScript & CWE-685 & 1 (100.00\%) & \makecell[l]{Function Call With Incorrect\\Number of Arguments} \\ 
    \bottomrule
    \end{tabularx}}
\end{table}

\bibliographystyle{splncs04}
\bibliography{cite}
\end{document}